\documentclass{aa}
\usepackage[varg]{txfonts}
\usepackage{graphicx}
\usepackage{natbib}
\usepackage{mathrsfs}
\usepackage{xcolor}

\def\deg{\ensuremath{\,{\rm deg}}\xspace}

\def\fh{\ensuremath{^{\mathrm h}}}
\def\fm{\ensuremath{^{\mathrm m}}}
\def\fs{\ensuremath{^{\mathrm s}}}
\def\fdg{\ensuremath{^\circ}}
\def\fmin{\ensuremath{^\prime}}
\def\fsec{\ensuremath{^{\prime\prime}}}
\newcommand\gdr[1]{\gaia~DR#1}
\newcommand\egdr[1]{\gaia~EDR#1}
\newcommand{\gaia}{Gaia\xspace}
\newcommand{\hip}{Hipparcos\xspace}
\newcommand{\orvara}{\texttt{orvara}\xspace}
\newcommand{\exofast}{\texttt{EXOFASTv2}\xspace}
\newcommand{\smw}{\ensuremath{S_{\rm MW}}\xspace}
\def\teff{\ensuremath{T_{\rm eff}}\xspace}
\def\logg{\ensuremath{\log g}\xspace}
\def\gmag{\ensuremath{G}\xspace}
\def\gbp{\ensuremath{G_{\rm BP}}\xspace}
\def\grp{\ensuremath{G_{\rm RP}}\xspace}
\def\parallax{\ensuremath{\varpi}\xspace}
\def\feh{\ensuremath{[\rm Fe/H]}\xspace}
\providecommand{\gcm}{\ensuremath{\,\rm{g}\,\rm{cm}^{-3}}\xspace}
\providecommand{\yr}{\ensuremath{\,\rm a}\xspace}
\providecommand{\pc}{\ensuremath{\,\rm pc}\xspace}
\providecommand{\Lsun}{\ensuremath{\,{\rm L}_{\odot}}\xspace}
\providecommand{\Msun}{\ensuremath{\,{\rm M}_{\odot}}\xspace}
\providecommand{\Rsun}{\ensuremath{\,{\rm R}_{\odot}}\xspace}
\providecommand{\Mjup}{\ensuremath{\,{\rm M}_{\rm Jup}}\xspace}
\providecommand{\mps}{\ensuremath{\,{\rm m\,s}^{-1}}\xspace}
\providecommand{\kmps}{\ensuremath{\,{\rm km\,s}^{-1}}\xspace}
\providecommand{\days}{\ensuremath{\,\rm d}\xspace}
\providecommand{\au}{\ensuremath{\,\rm au}\xspace}
\providecommand{\deg}{\ensuremath{\,\rm deg}\xspace}
\def\uprior{\ensuremath{\mathcal{U}}}
\providecommand{\sresinob}{\ensuremath{\sqrt{e_{\rm b}}\sin{\omega_{\rm b}}}\xspace}
\providecommand{\srecosob}{\ensuremath{\sqrt{e_{\rm b}}\cos{\omega_{\rm b}}}\xspace}
\def\mum{{\,$\mu$m}\xspace}
\def\mas{\,{mas}\xspace}
\def\masyr{{mas\,a$^{-1}$}\xspace}

\bibpunct{(}{)}{;}{a}{}{,}
\graphicspath{{images/}}

\title{A multi-technique detection of an eccentric giant planet around accelerating star HD 57625}

\author{D.~Barbato                  \inst{\ref{oapd}}
        \and D.~Mesa                \inst{\ref{oapd}}
        \and V.~D'Orazi             \inst{\ref{uniroma},\ref{oapd}}
        \and S.~Desidera            \inst{\ref{oapd}}
        \and A.~Ruggieri            \inst{\ref{oact}}
        \and J.~Farinato            \inst{\ref{oapd}}
        \and L.~Marafatto           \inst{\ref{oapd}}
        \and E.~Carolo              \inst{\ref{oapd}}
        \and D.~Vassallo            \inst{\ref{oapd}}
        \and S.~Ertel               \inst{\ref{steward},\ref{lbto}}
        \and J.~Hom                 \inst{\ref{steward}}
        \and R.M.~Anche             \inst{\ref{steward}}
        \and F.~Battaini            \inst{\ref{oapd}}
        \and A.~Becker              \inst{\ref{lbto}}
        \and M.~Bergomi             \inst{\ref{oapd}}
        \and F.~Biondi              \inst{\ref{mpe},\ref{oapd}}
        \and A.~Cardwell            \inst{\ref{lbto}}
        \and P.~Cerpelloni          \inst{\ref{oapd}}
        \and G.~Chauvin             \inst{\ref{mpia}}
        \and S.~Chinellato          \inst{\ref{oapd}}
        \and C.~Desgrange           \inst{\ref{mpia},\ref{grenoble}}
        \and S.~Di~Filippo          \inst{\ref{oapd}}
        \and M.~Dima                \inst{\ref{oapd}}
        \and T.S.~Gomes~Machado     \inst{\ref{oapd},\ref{unipd}}
        \and R.~Gratton             \inst{\ref{oapd}}
        \and D.~Greggio             \inst{\ref{oapd}}
        \and Th.~Henning            \inst{\ref{mpia}}
        \and M.~Kenworthy           \inst{\ref{leiden}}
        \and F.~Laudisio            \inst{\ref{oapd}}
        \and C.~Lazzoni             \inst{\ref{oapd}}
        \and J.~Leisenring          \inst{\ref{steward}}
        \and L.~Lessio              \inst{\ref{oapd}}
        \and A.~Lorenzetto          \inst{\ref{oapd}}
        \and L.~Mohr                \inst{\ref{mpia}}
        \and M.~Montoya             \inst{\ref{steward}}
        \and G.~Rodeghiero          \inst{\ref{oateramo}}
        \and J.~Patience            \inst{\ref{sese}}
        \and J.~Power               \inst{\ref{lbto}}
        \and D.~Ricci               \inst{\ref{oapd}}
        \and K.K.R.~Santhakumari    \inst{\ref{oapd}}
        \and A.~Sozzetti            \inst{\ref{oato}}
        \and G.~Umbriaco            \inst{\ref{unibo},\ref{oapd}}
        \and M.~Vega~Pallauta       \inst{\ref{uniroma}}
        \and V.~Viotto              \inst{\ref{oapd}}
        \and K.~Wagner              \inst{\ref{steward}}
        }
\institute{
            INAF – Osservatorio Astronomico di Padova, Vicolo dell’Osservatorio 5, I-35122, Padova, Italy \\ \email{domenico.barbato@inaf.it} \label{oapd}
            \and Department of Physics, University of Rome Tor Vergata, via della Ricerca Scientifica 1, 00133, Rome, Italy \label{uniroma}
            \and INAF – Osservatorio Astrofisico di Catania, Via S. Sofia 78, 95123 Catania, Italy \label{oact}
            \and Department of Astronomy and Steward Observatory, The University of Arizona, 933 North Cherry Ave, Tucson, AZ85721, USA \label{steward}
            \and Large Binocular Telescope Observatory, The University of Arizona, 933 North Cherry Ave, Tucson, AZ85721, USA \label{lbto}
            \and Max Planck Institute for extraterrestrial Physics, Gießenbachstraße 1, 85748 Garching bei München, Germany \label{mpe}
            \and Max Planck Institute for Astronomy,  Königstuhl 17, 69117 Heidelberg, Germany \label{mpia}
            \and Univ. Grenoble Alpes, CNRS, IPAG, F-38000 Grenoble, France \label{grenoble}
            \and Department of Physics and Astronomy, University of Padova, Vicolo dell’Osservatorio 3, 35122 Padova, Italy \label{unipd}
            \and Leiden Observatory, University of Leiden, PO Box 9513, 2300 RA Leiden, The Netherlands \label{leiden}
            \and INAF Osservatorio Astronomico d’Abruzzo, Via Mentore Maggini, I-64100 Teramo, Italy \label{oateramo}
            \and School of Earth and Space Exploration, Arizona State University, Tempe, AZ 85281, USA \label{sese}
            \and INAF – Osservatorio Astrofisico di Torino, Via Osservatorio 20, 10025, Pino Torinese, Italy \label{oato}
            \and Dipartimento di Fisica e Astronomia ”Augusto Righi” - Alma Mater Studiorum Università di Bologna, via Piero Gobetti 93/2 - 40129, Bologna, Italy \label{unibo}
            }
            
\date{Received <date> / Accepted <date>}
\abstract{The synergy between different detection methods is a key asset in exoplanetology, allowing for both precise characterization of detected exoplanets and robust constraints even in the case of non-detection. Recently, the interplay between imaging, radial velocities and astrometry has produced significant advancements in exoplanetary science.}
{We report a first result of an ongoing survey performed with SHARK-NIR, the new high-contrast near-infrared imaging camera at the Large Binocular Telescope, in parallel with LBTI/LMIRCam in order to detect planetary companions around stars with significant proper motion anomaly. In this work we focus on HD\,57625, a F8 star for which we determine a $4.8^{+3.7}_{-2.9}$ Ga age, exhibiting significant astrometric acceleration and for which archival radial velocities hint at the presence of a previously undetected massive long-period companion.}
{We analyse the imaging data we collected with SHARK-NIR and LMIRCam in synergy with the available public SOPHIE radial velocity time series and Hipparcos-Gaia proper motion anomaly. With this joint multi-technique analysis, we aim at characterizing the companion responsible for the astrometric and radial velocity signals.}
{The imaging observations result in a non-detection, indicating the companion to be in the substellar regime. This is confirmed by the synergic analysis of archival radial velocity and astrometric measurements resulting in the detection of HD\,57625\,b, a ${8.43}_{-0.91}^{+1.1}$\Mjup planetary companion with an orbital separation of ${5.70}_{-0.13}^{+0.14}$\au and ${0.52}_{-0.03}^{+0.04}$ eccentricity.}
{HD\,57625\,b joins the small but growing population of giant planets in outer orbits with true mass determination provided by the synergic usage of multiple detection methods, proving once again the importance of multi-technique analysis in providing robust characterization of planetary companions.}

\begin{document}
    \maketitle
  
    \section{Introduction} \label{sec:introduction}
        As the continuous growth of exoplanetology results in both new detections of planetary companions and refined and more in-depth characterization of known systems, the development of new and more precise instrumentation keeps fueling the exploration of all facets of exoplanetary science. As a result, the synergic interplay between the many discovery techniques available in the field becomes an increasingly key asset in furthering our search for exoplanetary systems. Among these various techniques, direct imaging is especially suited to detect giant companions on wide orbits ($\gtrsim$10\au) around young stars, and as such is able to provide an unparalleled view on the early stages of exoplanet formation and evolution processes. Some notable examples of exoplanetary systems discovered and characterized by imaging observations include HR\,8799 \citep{marois2008,marois2010}, GJ\,504 \citep{kuzuhara2013}, 51\,Eri \citep{macintosh2015}, PDS\,70 \citep{keppler2018,muller2018,haffert2019} and AF\,Lep \citep{mesa2023,franson2023,derosa2023}.
        \par Although successful, as every other discovery technique, the imaging method is characterized by drawbacks limiting its potential. Chiefly, imaging observations are highly time-consuming, requiring long exposures that can easily reach a few hours depending on target star characteristics, therefore impacting the scientific yield of observing nights. Additionally, the return of imaging surveys is intrinsically limited by the observed low occurrence rate of the wide-orbit massive companions for which the technique is ideally suited, often resulting in the need to observe hundreds of targets to successfully detect a low number of companions in the planetary mass regime \citep[see e.g.][]{galicher2016,nielsen2019,vigan2021}. Indeed, demographic studies suggest that the majority of giant planets have orbits with semi-major axes between 1 and 10\au, peaking for example at $\sim$3\au for M-dwarf host stars \citep[see e.g.][]{meyer2018}, putting a significant portion of giant exoplanets beyond the technical capabilities of current high-contrast imagers. As such, the relatively low imaging detection rate of planetary-mass companions represents a significant obstacle in bringing to full fruition the potential of imaging surveys.
        \par It is therefore clear that the usage of pre-selection criteria to identify the stellar targets more likely to host planetary companions within the reach of current imaging instruments is essential to enhance the scientific yield of imaging surveys. Such criteria, providing preliminary hints of the presence of planetary-mass companions and allowing for targeted observations, arise from the synergic usage of other detection techniques, again highlighting the increasing importance that the interplay between different techniques currently has in exoplanetology.
        \par A typical example of such pre-selection criteria is represented by the identification in radial velocity (RV) time series of either a full Keplerian orbital variation or trends hinting at the presence of undetected massive companions having an orbital period longer than the time span of the available RV observations, possibly ideal for imaging follow-up observations and robust detection. Successful examples of this strategy in the recent literature include the direct detection of brown dwarf (BD) GJ\,758\,B \citep{calissendorff2018} and exoplanets $\beta$\,Pic\,c \citep{nowak2020}, HD\,206893\,c \citep{hinkley2023} and $\epsilon$\,Ind\,Ab \citep{matthews2024} with imaging observations confirming previous RV-only results.
        \par Another successful pre-selection criterion focuses on the exploitation of the high-precision absolute astrometric catalogs produced by the Hipparcos \citep{vanleeuwen2007} and Gaia \citep{prusti2016} missions, in the latter case utilizing its second and third major data releases \citep[DR2 and DR3, see][]{brown2018,vallenari2023}. Specifically, a statistically significant difference between the long-term proper motion vector of a star common to the two catalogs and the short-term Gaia (and Hipparcos) measurements, typically referred to as proper motion anomaly (PMa) will indicate the presence of a perturbing companion. The recent release of \hip-\gaia astrometric acceleration catalogs \citep{brandt2021hgca,kervella2022} has fueled the usage of significant PMa as pre-selection tool for imaging surveys and resulted in the detection of brown dwarfs (BDs) companions \citep[see e.g.][]{calissendorff2018,currie2020,bonavita2022} and is recently starting to show its exoplanetary potential in the detection of AF\,Lep\,b \citep[see e.g.][]{mesa2023,derosa2023,franson2023}.
        \par In both cases, imaging observations can either successfully detect the planetary companion responsible for the RV trend or the significant PMa or, in the case of a non-detection, provide robust constraints on both companion mass and orbital separation based on the detection capabilities of the specific instrument \citep[see e.g.][]{mesa2022}. As such, even a non-detection can advance the characterization of candidate planetary companions and fuel follow-up observations, once more proving the opportunities represented by the synergic usage of multi-technique analysis.
        \par As such, the observation of a sample of stars exhibiting significant astrometric acceleration is a key scientific program selected for the early science validation phase of SHARK-NIR \citep{farinato2022,farinato2023,marafatto2022}, a high-contrast camera operating in the Y, J and H near-infrared bands that has recently been installed on the left arm of the Large Binocular Telescope (LBT) in Arizona. The instrument is designed to fully take advantage of the extreme adaptive optics (AO) correction performed by the LBT AO system SOUL \citep[Single Conjugated Adaptive Optics Upgrade for LBT, see][]{pinna2016,pinna2023}. Indeed, detailed simulations have shown that SHARK-NIR is able to achieve in good weather Strehl ratios up to more than 90\% and contrast levels down to $10^{-6}$ at separations of 300-500 mas \citep{carolo2018}.  The instrument's location at the LBT common center-bent Gregorian focus allows for synergic simultaneous high-contrast observations with SHARK-VIS \citep[operating in visible light, see][]{pedichini2022} and LBTI/LMIRCam \citep[operating in J to M infrared bands, see][]{skrutskie2010,leisenring2012}. Installed on LBT in October 2022, SHARK-NIR has afterwards completed its commissioning phase and is now undertaking its early science validation runs \citep[as detailed in][]{barbato2024} directly addressing its scientific goals, namely the detection and characterization of giant extrasolar planets, disks and jets around young stars, active galactic nuclei and Solar System small bodies.
        \par In this work we present the first results from the accelerating stars survey to be further presented in \cite{mesa2024}, performed with SHARK-NIR during its first early science runs, more specifically focusing on the observations undertaken in February 2024 on star HD\,57625 (HIP\,36014). This star was selected as part of SHARK-NIR early science observations both by virtue of its significant PMa reported in \cite{kervella2022}, with a signal-to-noise ratio (SNR) of 11.04, and archival SOPHIE RV time series showing a clear long-term variation, both characteristics hinting at the presence of a previously undetected long-period giant planetary companion that we present and characterize for the first time via a multi-technique analysis.
        \par This paper is organized as follows: in Sect. \ref{sec:star-parameters} we provide an updated description of the target star physical characteristic, before moving on to describe the SHARK-NIR and LMIRCam observations and data analysis in Sect. \ref{sec:imaging}, the analysis of the archival radial velocity data in Sect. \ref{sec:rv} and of the proper motion anomaly measurements in Sect. \ref{sec:pma}. Finally, we perform a multi-technique detection completeness characterization of the target star in Sect. \ref{sec:completeness}, before concluding and discussing our results and future perspectives in Sect. \ref{sec:conclusions}.
        
    \section{HD\,57625 stellar characteristics}    \label{sec:star-parameters}
        \begin{table}
          \caption{Stellar parameters for HD\,57625.}       \label{table:star-parameters}
            \centering
            \begin{tabular}{l c}
                \hline\hline
                    \multicolumn{2}{c}{HD\,57625}\\
                \hline
                    HIP                                     & 36014\\[3pt]
                    Gaia DR3                                & 989016898234332416\\[3pt]
                    $\alpha$ (J2000)\tablefootmark{a}       & 7\fh25\fm18.03\fs\\[3pt]
                    $\delta$ (J2000)\tablefootmark{a}       & +56\fdg34\fmin8.96\fsec\\[3pt]
                    \parallax (\mas)\tablefootmark{a}       & $22.5622\pm0.0263$\\[3pt]
                    $\mu_\alpha$ (\masyr)\tablefootmark{a}  & $-55.181\pm0.020$\\[3pt]
                    $\mu_\delta$ (\masyr)\tablefootmark{a}  & $-59.102\pm0.016$\\[3pt]
                    $\Delta\mu_\alpha$ (\masyr)\tablefootmark{b} & $0.340\pm0.026$ \\[3pt]
                    $\Delta\mu_\delta$ (\masyr)\tablefootmark{b} & $-0.141\pm0.021$ \\[3pt]
                    d$V_{\rm t}$ (\mps)\tablefootmark{b} & $77.34\pm7.01$ \\[3pt]
                    $RV_{\rm sys}$ (\kmps)\tablefootmark{c} & $4.673\pm0.0024$\\[3pt]
                    $B_T$ (mag)\tablefootmark{d}            & $8.41\pm0.01$\\[3pt]
                    $V_T$ (mag)\tablefootmark{d}            & $7.75\pm0.01$\\[3pt]
                    $B$ (mag)\tablefootmark{e}              & $8.41\pm0.01$\\[3pt]
                    $V$ (mag)\tablefootmark{e}              & $7.75\pm0.01$\\[3pt]
                    $J$ (mag)\tablefootmark{e}              & $6.581\pm0.019$\\[3pt]
                    $H$ (mag)\tablefootmark{e}              & $6.364\pm0.038$\\[3pt]
                    $K$ (mag)\tablefootmark{e}              & $6.276\pm0.017$\\[3pt]
                    WISE1 (mag)\tablefootmark{g}            & $6.214\pm0.067$\\[3pt]
                    WISE2 (mag)\tablefootmark{g}            & $6.214\pm0.023$\\[3pt]
                    WISE3 (mag)\tablefootmark{g}            & $6.293\pm0.015$\\[3pt]
                    WISE4 (mag)\tablefootmark{g}            & $6.303\pm0.055$\\[3pt]
                    \gmag (mag)\tablefootmark{c}            & $7.555\pm0.003$\\[3pt]
                    \gbp (mag)\tablefootmark{c}             & $7.842\pm0.003$\\[3pt]
                    \grp (mag)\tablefootmark{c}             & $7.098\pm0.004$\\[3pt]
                    Spectral Type\tablefootmark{h}          & F8\\[3pt]
                    $M_\star$ (\Msun)\tablefootmark{i}      & $1.040^{+0.078}_{-0.080}$\\[3pt]
                    $R_\star$ (\Rsun)\tablefootmark{i}      & $1.126^{+0.046}_{-0.040}$\\[3pt]
                    $\rho_\star$ (\gcm)\tablefootmark{i}    & $1.02^{+0.15}_{-0.14}$\\[3pt]
                    $L_\star$ (\Lsun)\tablefootmark{i}      & $1.466^{+0.13}_{-0.093}$\\[3pt]
                    Age (Ga)\tablefootmark{i}              & $4.8^{+3.7}_{-2.9}$\\[3pt]
                    \teff (K)\tablefootmark{j}              & $5961.78\pm43.33$\\[3pt]
                    \logg (cgs)\tablefootmark{j}            & $4.47\pm0.08$\\[3pt]
                    \feh\tablefootmark{j}                   & $-0.04\pm0.02$\\[3pt]
                    $v\sin{i}$ (\kmps)\tablefootmark{k}     & $1.8$\\[3pt]
                    \smw\tablefootmark{l}                   &$0.1898\pm0.0367$\\[3pt]
                    H$\alpha$\tablefootmark{l}              &$0.1294\pm0.0060$\\[3pt]
                    Na I\tablefootmark{l}                   &$0.4962\pm0.0157$\\[3pt]
                    $P_{\rm{rot,\,Noyes}}$ (d)\tablefootmark{l}&$22.25\pm4.23$\\[3pt]
                    $P_{\rm{rot,\,Mamajek}}$ (d)\tablefootmark{l}&$22.47\pm4.96$\\[3pt]
                \hline
            \end{tabular}
            \tablefoot{
                          \tablefoottext{a}{retrieved from Gaia Data Release 3 \citep{vallenari2023}}
                          \tablefoottext{b}{retrieved from \cite{kervella2022}}
                          \tablefoottext{c}{retrieved from Gaia Data Release 2 \citep{soubiran2018}}
                          \tablefoottext{d}{retrieved from \cite{hog2000}}
                          \tablefoottext{e}{retrieved from \cite{zacharias2012}}
                          \tablefoottext{f}{retrieved from \cite{cutri2003}}
                          \tablefoottext{g}{retrieved from \cite{cutri2014}}
                          \tablefoottext{h}{retrieved from \cite{cannon1993}}
                          \tablefoottext{i}{obtained from the SED fitting discussed in Sect. \ref{sec:star-parameters}}
                          \tablefoottext{j}{obtained from the spectroscopic analysis discussed in Sect. \ref{sec:star-parameters}}
                          \tablefoottext{k}{retrieved from \cite{glebocki2005}}
                          \tablefoottext{l}{obtained from the SOPHIE spectra analysis discussed in Sect. \ref{sec:star-parameters}}
                        }
        \end{table}
        \begin{table*}
            \caption{SOPHIE and SOPHIE+ measurements for HD\,57625}    \label{table:rvdata}
            \centering
            \begin{tabular}{l c c c c c c c}
                \hline\hline
                    BJD & RV & FWHM & BIS & \smw & H$\alpha$ & Na~{\sc i}\\
                        & (\kmps) & (\kmps) & (\kmps)  & & & \\[3pt]
                \hline
                    $2454051.70$ & $4.7913\pm0.0027$ & $7.7459$ & $0.0000$ & $0.3768\pm0.0021$& $0.1490\pm0.0012$& $0.5103\pm0.0023$\\
                    $2454053.66$ & $4.7942\pm0.0027$ & $7.7566$ & $0.0000$ & $0.3277\pm0.0016$& $0.1416\pm0.0009$& $0.4844\pm0.0016$\\
                    $2454872.47$ & $4.6845\pm0.0027$ & $7.7541$ & $0.0038$ & $0.1900\pm0.0028$& $0.1373\pm0.0014$& $0.4900\pm0.0028$\\
                    $2454893.44$ & $4.6813\pm0.0026$ & $7.7820$ & $-0.0048$ & $0.2099\pm0.0027$& $0.1441\pm0.0014$& $0.5015\pm0.0028$\\
                    $2454894.36$ & $4.6813\pm0.0026$ & $7.7816$ & $0.0078$ & $0.2524\pm0.0031$& $0.1386\pm0.0014$& $0.5044\pm0.0027$\\
                    ... & ... & ... & ... & ... & ... & ... \\
                \hline
            \end{tabular}
            \tablefoot{Data are available at the CDS and on the SOPHIE archive. A portion is shown here for guidance regarding its form and content.}
        \end{table*}
        HD\,57625 (HIP\,36014) is a F8 star distant 44.24\pc from the Sun and estimated to be 6.29 Ga old \citep{vallenari2023} with a reported mass of $1.16\pm0.06$\Msun \citep{kervella2022}. As mentioned in Sect.~\ref{sec:introduction}, the astrometric acceleration catalog presented in \citep{kervella2022} reports a strong PMa signal having a SNR of 11.04, hinting at the likely presence of an unseen massive companion. 
       \par  HD\,57625 is part of a wide binary system, with a fainter stellar companion (2MASS J07251770+5634002, Gaia DR3 989016790859521280, \gmag=15.17\,mag) located at 10$^{\prime\prime}$ corresponding to a projected separation of $\sim$440\au. The physical association is supported by the Gaia astrometric parameters.\footnote{We additionally note that HD\,57625 is included in the current version of the Washington Double Star (WDS) catalog as a triple system, but both the listed components, namely HD\,57667 (WDS J07254+5633A) and  2MASS J07251750+5633363 (WDS J07254+5633C) are clearly ruled out as physical companions by the largely discrepant astrometric parameters in Gaia DR3.} A mass of $\sim$0.2\Msun is estimated for the companion from the main sequence relationships by \cite{pecaut2013}\footnote{Updated version (2022.04.16) at \url{https://www.pas.rochester.edu/~emamajek/EEM_dwarf_UBVIJHK_colors_Teff.txt}}. We note that the observed PMa is not compatible with this distant low-mass stellar companion, as we will discuss in Sect.~\ref{sec:pma}.
       \par For further characterization, we exploited archival spectra taken with SOPHIE as described in Sect. \ref{sec:rv} to derive atmospheric parameters and metallicity. We employed {\tt LOTUS} (non-LTE Optimization Tool Utilized for the derivation of atmospheric Stellar parameters) by \cite{li2023}\footnote{\url{https://lotus-nlte.readthedocs.io/en/latest/}} adopting an optimised linelist for solar analogues, which is available upon request. This tool calculates parameters using the equivalent width (EW) method for Fe I and II lines, incorporating a generalized curve of growth approach to account for EW dependencies on the corresponding atmospheric stellar parameters. A global differential evolution optimization algorithm is applied to extract the fundamental parameters. Furthermore, {\tt LOTUS} provides precise uncertainties for each stellar parameter through a Markov Chain Monte Carlo algorithm. The resulting spectroscopic effective temperature (\teff) is 5962$\pm$43 K, the stellar gravity is $\log g$ 4.47$\pm$0.08 dex, the microturbulent velocity is V$_{\rm mic}$=1.00$\pm$0.03\kmps, and the iron abundance [Fe/H] is $-0.04\pm$0.02 (random errors). These results are consistent, within observational uncertainties, with the available estimates in the literature.
       \par In order to have an updated and homogeneous characterisation of the physical properties of HD\,57625 for our analysis, we fit its stellar spectral energy distribution (SED) using the MESA Isochrones and Stellar Tracks (MIST) \citep{dotter2016,choi2016} via the \texttt{IDL} suite \exofast \citep{eastman2019}. With this method, the stellar parameters are simultaneously constrained by the SED and the MIST isochrones, since the SED primarily constrains the stellar radius $R_\star$ and effective temperature \teff, while a penalty for straying from the MIST evolutionary tracks ensures that the resulting star is physical in nature \citep[see][for more details on the method]{eastman2019}. We considered all available archival magnitudes from Tycho $B_T$ and $V_T$ bands \citep{hog2000}, Johnson’s $B$, $V$ and 2MASS $J$, $H$, and $K$ bands from the UCAC4 catalog \citep{zacharias2012,cutri2003}, WISE bands \citep{cutri2014}, and \gaia \gmag, \gbp, and \grp bands \citep{vallenari2023}, imposing Gaussian priors on the star's effective temperature \teff, and metallicity \feh based on their respective values we obtain from the activity indexes analysis detailed in the previous paragraph, as well as on the stellar parallax \parallax based on \gaia~DR3 astrometric measurement \citep{vallenari2023}. We list in Table \ref{table:star-parameters} the resulting best-fit stellar parameters, as well as the other parameters resulting from all analysis detailed in this Section.
       \par The resulting age from isochrones is in general agreement with the evidences from other indicators, such as the low activity inferred from the SOPHIE spectra discussed in the next paragraph, X-ray non detection in Rosat All Sky Survey \citep{voges1999,voges2000}, slow rotation \citep{glebocki2005}, thin disk kinematics \citep{smart2021}. Therefore, we adopt the isochrone age and corresponding stellar mass in the rest of this work.
       \par To determine the activity level of HD\,57625 and check whether the detected RV signal (see Sect. \ref{sec:rv}) is related to an activity cycle, we use the Python tool \texttt{ACTIN2}\footnote{\url{https://actin2.readthedocs.io/en/latest/index.html}} \citep{gomesdasilva2018, gomesdasilva2021} to extract activity indicators from the SOPHIE spectra. In particular, we used the following indicators: emission from the Ca II H \& K lines (\smw), H$\alpha$, and Na I D2 line, whose median values and corresponding $1\sigma$ error bars are shown in Table~\ref{table:star-parameters}. We plot in Fig~\ref{fig:rvgls} the resulting activity indexes time series, as well as the RV, full width at half maximum (FWHM) and bisector inverse span (BIS) we analyse in detail in Sect.~\ref{sec:rv}. We produce generalized Lomb-Scargle (GLS) periodograms for each activity index time series, finding for \smw a long-period oscillation (>5000\days), while for H$\alpha$ we found a very significant (FAP $\ll0.001\%$) peak at 11.35\days, as well as another long-period (>5000\days) trend. Interestingly, the 11.35\days peak is exactly half the rotation period of the star derived from the \smw time series using the equations by \cite{noyes1984} and \cite{mamajek2008}. In particular, we obtained $P_{\rm{rot,\,Noyes}} = 22.25 \pm 4.23$\days and $P_{\rm{rot,\,Mamajek}} = 22.47 \pm 4.96$\days, in perfect agreement with each other. For the Na I, we find a low-significance (FAP $= 2.6\%$) principal peak at 4\days, indicating no strong signal in this time series. From this analysis, we conclude that HD\,57625 lacks any strong activity, excluding possible long-period cycles inferred from the \smw and H$\alpha$ time series. We additionally note that all activity indexes analysed appear to be uncorrelated to the observed large RV variation, with the largest Pearson correlation coefficient $\rho$ being that of \smw hinting at a moderate correlation ($\rho=0.49$) which we will further address in Sect.~\ref{sec:rv}.
        
    \section{Imaging observations and analysis}    \label{sec:imaging}
        \begin{figure*}
          \includegraphics[width=\linewidth]{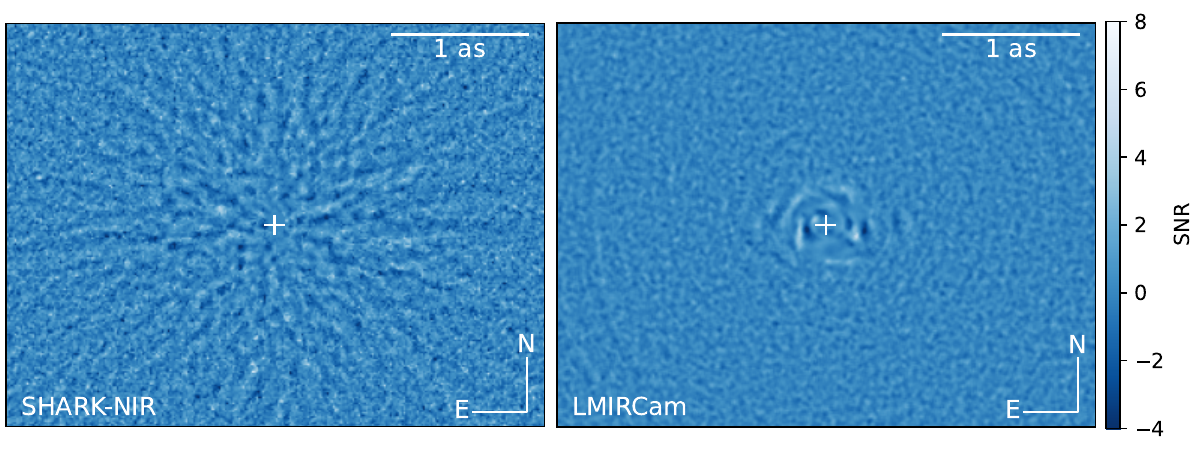}
          \caption{Final SNR map of the inner region around HD\,57625 obtained using the SHARK-NIR H-band (\textit{left panel}) and the LMIRCam L$^\prime$-band data (\textit{right panel}), a white cross marking the position of the central star. In both cases the images were obtained using a PCA technique subtracting 10 principal components.}
          \label{fig:imasnr}
        \end{figure*}
        \begin{figure}
          \includegraphics[width=\linewidth]{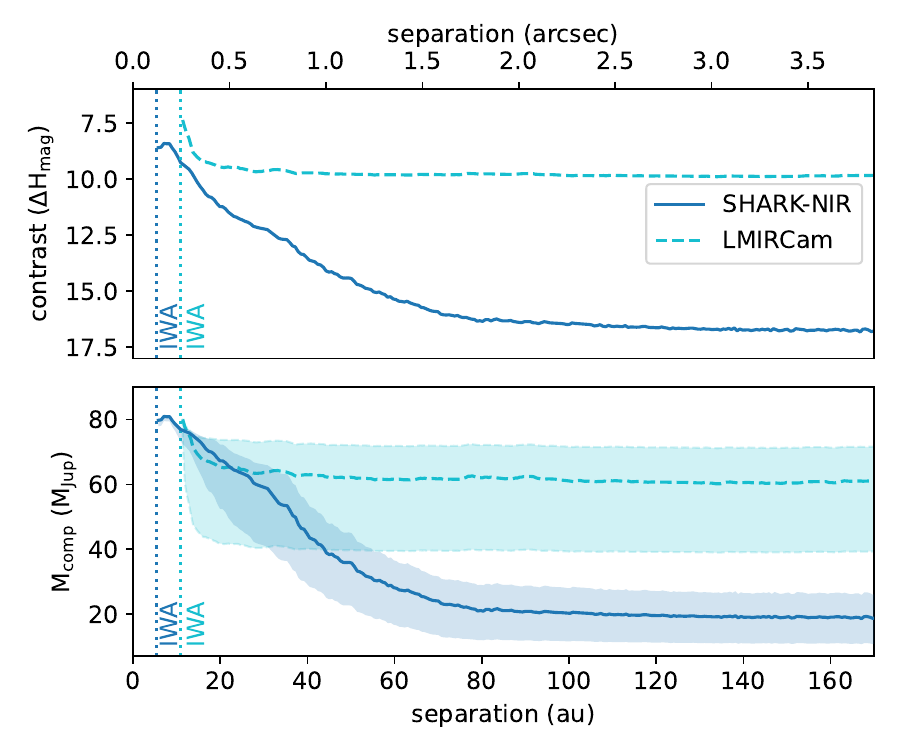}
          \caption{\textit{Top panel:} contrast curve for the imaging observation of HD\,57625 conducted on the night of February 24th 2024 UT. \textit{Bottom panel}: mass limits derived using the AMES-COND models, the thick curve corresponding to the nominal 4.8 Ga stellar age, the shaded region corresponding to the age uncertainty. In both panels, the blue solid curves refer to the SHARK-NIR H-band observations and the cyan dashed lines refer to the LMIRCam L$^\prime$-band observation, with similarly colour-coded vertical dotted lines indicating the Inner Working Angle (IWA) of each instrument.}
          \label{fig:contrast}
        \end{figure}
        We observed HD\,57625 with SHARK-NIR on the night of February 24th 2024 UT using the broadband H filter (central wavelength at 1.6\mum and 0.218\mum bandwidth) and the instrument's Gaussian coronagraph that has an inner Working Angle (IWA) of 120\mas, which at the 44.24\pc distance of HD\,57625 corresponds to a 5.30\au separation. We obtained a total of 208 frames each with a 18.7257\,s detector integration time (DIT) for a total exposure time of $\sim$3895\,s ($\sim$65 minutes). The observations were performed in pupil stabilized mode in order to implement angular differential imaging \citep[ADI, see e.g.][]{marois2006}, allowing a total rotation of the field of view (FOV) of 40.58\deg. During the observations the seeing varied between 0.94$^{\prime\prime}$ and 2.19$^{\prime\prime}$. We also took an image with the stellar PSF unocculted by the coronagraph both before and after the main coronagraphic observations, to be able to correctly estimate the contrast on the scientific image. These frames were taken introducing in the optical path a neutral density filter (ND3) to avoid the detector saturation.
        \par In addition to the SHARK-NIR observations we simultaneously observed our target with LBTI/LMIRCam  \citep{skrutskie2010,leisenring2012} on the right arm of LBT. These observations were performed in the L$^\prime$ spectral band (central wavelength at 3.7\mum and 0.58\mum bandwidth) using the vector-apodizing phase plate coronagraph \citep[vAPP,][]{doelman2021}, with an IWA of 246\mas corresponding for HD\,57625 to a 10.88\au separation. In this case we obtained 3200 frames, each of them with a DIT of 1.51085\,s for a total exposure time of 4834.72\,s (80.58 minutes). The LMIRCam observations were also performed in pupil stabilized mode with a total rotation of the FOV of 43.38\deg. 
        \par Because of the non-destructive charge transfer capability of the SHARK-NIR scientific detector, each individual exposure is a ramp, i.e. a set of reads taken at uniform time intervals. The raw ramps were first reduced using a Python pipeline specifically developed for the SHARK-NIR detector, performing reference pixel correction, bias subtraction, and linearity correction. Finally, the software collapses the ramps into single frames using an up-the-ramp sampling algorithm.
        \par The SHARK-NIR data were then post-processed using the pipeline custom designed for the instrument and written in Python. First, we subtracted the dark and divided by the flat-field images. For our analysis, we meticulously fine-centered each frame by utilizing the four symmetrical spots created by the SHARK-NIR internal deformable mirror. Additionally, we conducted a selection process to identify and exclude any low-quality frames such as those where the star was not adequately masked by the coronagraph. Finally, we applied a post-processing procedure based on the ADI method and exploiting the principal components analysis \citep[PCA;][]{soummer2012,amara2012} algorithm. 
        \par Similarly, each LMIRCam exposure is a ramp composed in this case of two frames taken at different intermediate exposure times, the first one with a time equal to just 13.7 ms and the last one corresponding to the total exposure time. As a first step, we subtracted the first image of the ramp data cube from the last one to remove the bias from the science data. The data were taken with two different nodding positions called A and B. We then subtracted from each image a median of the 100 closest images in time from the opposite nod position, in order to subtract thermal background flux from all frames obtained at MIR wavelengths without needing to obtain sky frames. Finally, we applied a PCA-based post-processing similar to what was done for the SHARK-NIR data.
        \par The final images obtained for both instruments from the procedure described above are displayed in Fig.~\ref{fig:imasnr}.  No source with SNR$>$5 is detected on either image, allowing us to conclude that no companion has been detected orbiting around HD\,57625 during our observations. However, as discussed in Sect.~\ref{sec:introduction} even a non-detection in imaging data can be successfully used to provide robust limits on the unseen companion producing the PMa and RV signals.
        \par In Fig.~\ref{fig:contrast} we show the SHARK-NIR and LMIRCam observation contrast curve as a function of separation as well as the respective mass limits derived using the AMES-COND models \citep{allard2003} assuming the stellar age range of $4.8^{+3.7}_{-2.9}$ Ga derived in Sect.~\ref{sec:star-parameters}. From these mass limits, it is clear that while any planetary companion of HD\,57625 would not have been detected, both SHARK-NIR and LMIRCam would have been able to detect any stellar companion at all separations above the aforementioned respective coronagraphic limits of 5.3\au and 10.88\au, as well as very wide and massive brown dwarfs. Therefore, the non-detection produced by our parallel SHARK-NIR and LMIRCam imaging observations in H and L$^\prime$ infrared bands allows to confirm the substellar and likely planetary nature of the companion producing the PMa and RV signal of HD\,57625.
        
    \section{Radial velocity analysis}    \label{sec:rv}
        \begin{figure}
            \includegraphics[width=\linewidth]{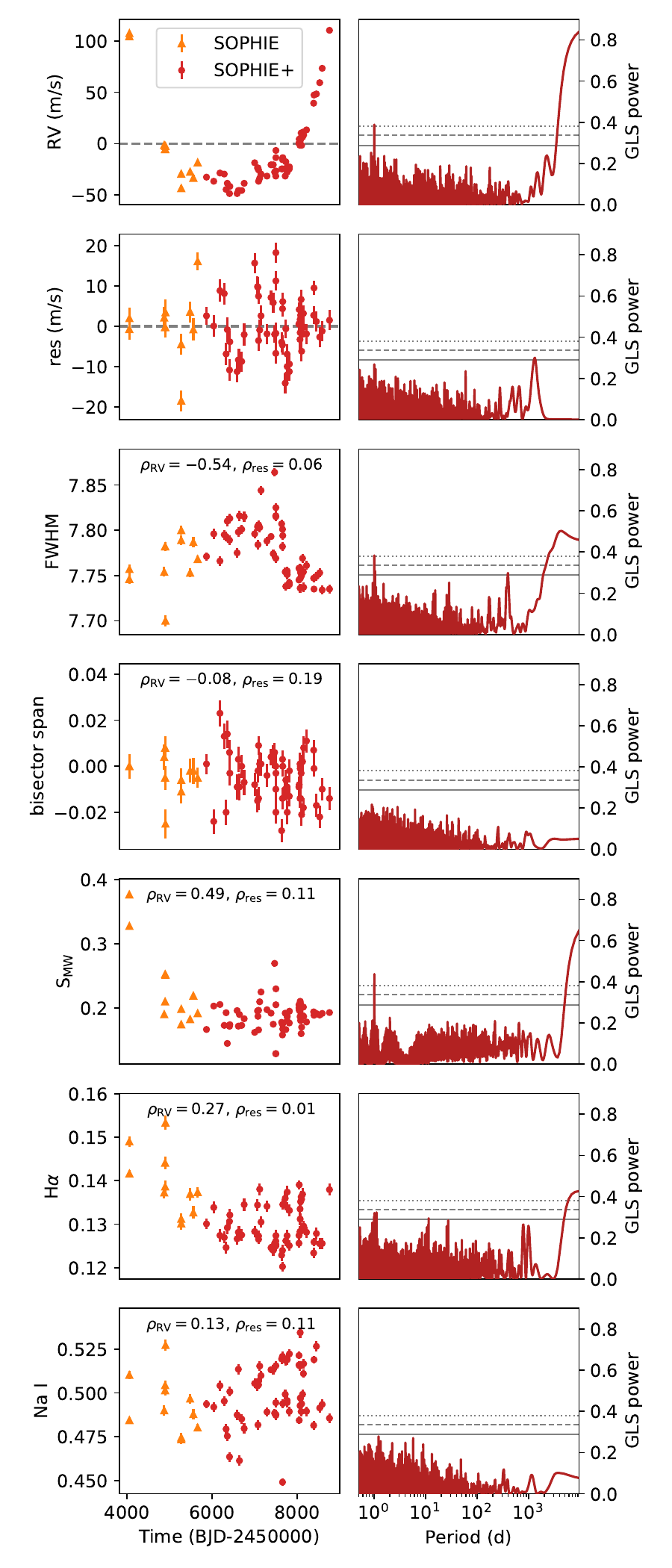}
            \caption{\textit{Left panels}: time series for the SOPHIE (orange triangles) and SOPHIE+ (red circles) radial velocity data of HD\,57625, residual time series obtained after removing the detected Keplerian signal and activity indexes. For the activity indexes time series, the Pearson correlation coefficient with both original and residual RVs are noted. \textit{Right panels}: corresponding generalised Lomb-Scargle periodograms of the time series, with horizontal solid, dashed and dotted lines marking the 10\%, 1\% and 0.1\% FAP thresholds.}
            \label{fig:rvgls}
        \end{figure}
        \begin{figure}
            \includegraphics[width=\linewidth]{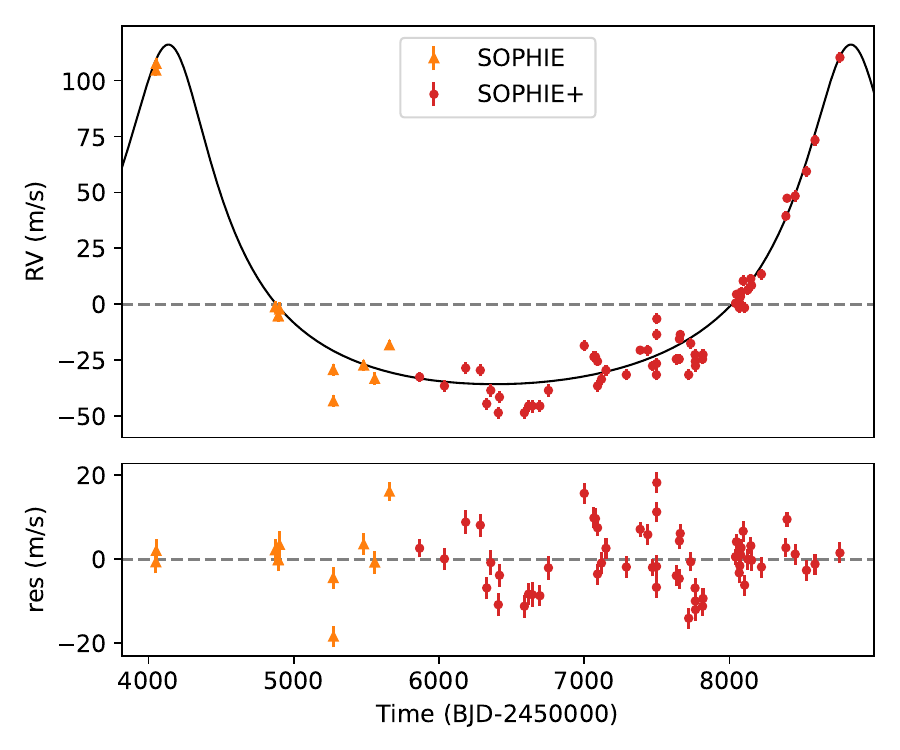}
            \caption{Radial velocity orbital fit for planet HD\,57625\,b. \textit{Top panel}: best-fit single-Keplerian solution is shown as a black curve over the SOPHIE (orange triangles) and SOPHIE+ (red circles) archival data. \textit{Bottom panel}: post-fit residual radial velocities.}
            \label{fig:rvfit}
        \end{figure}
        As mentioned in Sect.~\ref{sec:introduction}, the archival radial velocity time series available for HD\,57625 are characterized by the presence of a clear long-term trend hinting at the presence of a wide-orbit massive companion. Specifically, in this Section we focus on the publicly available RV measurements collected by the spectrograph SOPHIE \citep{perruchot2008}, mounted on the 1.93 m telescope at the Haute-Provence Observatory in France. We additionally note that in June 2011 the fibre link of the instrument was upgraded \citep{bouchy2013}, introducing an offset between the data collected before and after this date. In the following we therefore treat the RV measurements taken for HD\,57625 before and after the fibre upgrade as independent datasets named as SOPHIE and SOPHIE+, respectively.
        \par We retrieve a total of 70 publicly available radial velocity measurements (11 before the fibre upgrade and 59 after) from the SOPHIE archive\footnote{\url{http://atlas.obs-hp.fr/sophie}} \citep{moultaka2004}, spanning from November 12th 2006 to October 2nd 2019 and therefore providing an observational baseline of 4708\days with a median RV uncertainty of 2.5\mps. 
        In Fig.~\ref{fig:rvgls} we show the time series and GLS periodograms for the SOPHIE and SOPHIE+ RVs, the one-Keplerian solution residuals and the activity indexes derived and discussed in Sect.~\ref{sec:star-parameters}; all time series are also listed in Table~\ref{table:rvdata}. The RV time series exhibits a peak-to-valley variation of 151\mps and a clear Keplerian behaviour hinting at the presence of a moderate-to-high eccentricity companion with a GLS-derived period longer than 4500\days. While it must be noted that the available public data fail to fully cover the entirety of the RV oscillation, the current coverage allow for a significant portion of the oscillation to be detected and analysed. We note, as mentioned in Sect.~\ref{sec:star-parameters}, that no activity index shows significant correlation with the RV time series, with all indexes having Pearson correlation coefficient $\left|\rho\right|<0.27$ with the exceptions of FWHM ($\rho=-0.54$) and \smw ($\rho=0.49$) having moderate correlation. However, we note that both time series and periodogram of these indexes hint at different periodicities than the $>$4500\days observed in the RVs, with the FWHM periodogram peaking at $\sim$3900\days and the \smw variation not having gone through a full cycle over the observational time span of 4708\days and therefore strongly suggesting a periodicity much longer than that observed in the RVs.
        \par Although we noted in Sect.~\ref{sec:star-parameters} the presence of a $\sim$0.2\Msun stellar companion at a separation of 10$^{\prime\prime}$, we exclude a binary origin for the observed RV variation. With a projected orbital separation of $\sim440$\au, the orbital period of the secondary around the primary would be of $\sim8.5\cdot10^3$\yr, an exceedingly long period compared to the observed RV oscillation. Additionally, we can follow \cite{torres1999} and estimate the acceleration $d(RV)/dt$ caused by the secondary on the primary as:
            \begin{equation}        \label{eq:binary-dvdt}
                \frac{d(RV)}{dt}= G \frac{M_B}{a^2 (1-e)} \frac{(1+\cos{\nu})\sin{(\nu+\omega)}\sin{i}}{(1+\cos{E})(1-e\cos{E})}
            \end{equation}
        being  $a=a_A (M_A+M_B)/M_B$ the semi-major axis of the relative orbit, $\nu$ is the true anomaly, $i$ is the mutual inclination, and $E$ is the eccentric anomaly. As we lack any estimate of the binary system orbital elements except the aforementioned separation and mass estimates, we generated $10^5$ possible combinations of orbital elements $(e,v,\omega,E,i),$. We obtain a maximum possible variation of only $\sim$0.19\mps over the 12.5\yr SOPHIE baseline, an exceedingly lower value than the 151\mps variation observed in the archival data.
        \begin{table}
            \caption{RV-only best-fit orbital solution for HD\,57625\,b.}    \label{table:rvfit}
            \centering
            \begin{tabular}{l c c}
                \hline\hline
                    Parameter & Priors & Best-fit values\\
                \hline
                    $P_{\rm b}$ (\days) & \uprior$(2000,10000)$ & $4851_{-178}^{+413}$ \\[3pt]
                    $K_{\rm b}$ (\mps) & \uprior$(0,+\infty)$ & $78.94_{-5.85}^{+19.87}$ \\[3pt]
                    \sresinob & \uprior$(-1,1)$ & $0.006_{-0.058}^{+0.072}$ \\[3pt]
                    \srecosob & \uprior$(-1,1)$ & $0.723_{-0.028}^{+0.026}$ \\[3pt]
                    $\lambda_0$ (deg) & \uprior$(0,360)$ & $47.82_{-17.68}^{+9.44}$ \\[3pt]
                    $\omega_{\rm b}$ (deg) & derived & $0.46_{-4.70}^{+5.54}$ \\[3pt]
                    $e_{\rm b}$ & derived & $0.52\pm0.04$ \\[3pt]
                    $M_{\rm b}\sin{i}$ (\Mjup) & derived & $5.79_{-0.63}^{+1.71}$ \\[3pt]
                    $a_{\rm b}$ (au) & derived & $5.71_{-0.21}^{+0.33}$ \\[3pt]
                    $\gamma_{\rm SOPHIE}$ (\mps) & \uprior$(-\infty,+\infty)$ & $14.41_{-5.37}^{+9.55}$ \\[3pt]
                    $\gamma_{\rm SOPHIE+}$ (\mps) & \uprior$(-\infty,+\infty)$ & $22.62_{-5.17}^{+11.09}$ \\[3pt]
                    $j_{\rm SOPHIE}$ (\mps) & \uprior$(0,+\infty)$ & $9.55_{-2.19}^{+3.38}$ \\[3pt]
                    $j_{\rm SOPHIE+}$ (\mps) & \uprior$(0,+\infty)$ & $6.85_{-0.70}^{+0.81}$ \\[3pt]
                \hline
            \end{tabular}
        \end{table}
        \par We therefore search for the best-fit orbital solution using the Python tool \texttt{PyORBIT}\footnote{\url{https://github.com/LucaMalavolta/PyORBIT}} \citep{malavolta2016,malavolta2018}, a package for the Markov chain Monte Carlo (MCMC) modeling of RV and activity indexes time series based on the optimization algorithm \texttt{PyDE} \citep{storn1997} and the MCMC sampler \texttt{emcee} \citep{foreman2013}. The nine free parameters we fit for are orbital period $P$, semi-amplitude $K$, mean longitude $\lambda_0$, \sresinob, \srecosob, and finally a zero-point radial velocity term $\gamma$ and an uncorrelated stellar jitter term $j$ each for SOPHIE and SOPHIE+ datasets.
        \par We find a best-fit Keplerian solution (see Fig.~\ref{fig:rvfit} and Table~\ref{table:rvfit}) with orbital period $4851_{-178}^{+413}$\days, semi-amplitude $78.94_{-5.85}^{+19.87}$\mps and eccentricity $0.52\pm0.04$. Using the host star mass value of $1.040^{+0.078}_{-0.080}$\Msun we obtained in Sect.~\ref{sec:star-parameters}, we derive a companion minimum mass of $5.79_{-0.63}^{+1.71}$\Mjup and semi-major axis of $5.71_{-0.21}^{+0.33}$\au. We immediately note that both $P$ and especially $K$ of the detected companion upper limits are somewhat unconstrained, with upper relative uncertainties of $8.5\%$ and $25\%$, a clear result of the aforementioned incomplete orbital coverage of the available SOPHIE public data. Indeed, not only the best-fit orbital period is close to the observational baseline of 4708\days, but the available RV measurements are characterized by a lack of full sampling over the maximum portion of the companion-induced oscillation, therefore limiting a full characterization of some orbital elements of the planetary companion with the exclusive use of currently available RV observations. Finally, the residual time series has a weighted root mean square (w.r.m.s.) of $7$\mps, has no significant correlation with any of the activity indexes analysed ($\left|\rho\right|<0.2$) and exhibits a non-significant principal peak at $\sim$1300\days with a $6\%$ FAP in its GLS periodogram (see second row of Fig.~\ref{fig:rvgls}).
        \par As such, the HD\,57625 system calls for further follow-up RV observations to fully constrain the orbital parameters of planet b. However, as we will show in the following Sections, the simultaneous usage of astrometric and imaging measurements can provide additional constrains to the RV-only system characterization.
        
    \section{Astrometry analysis}    \label{sec:pma}
        \begin{figure}
          \includegraphics[width=\linewidth]{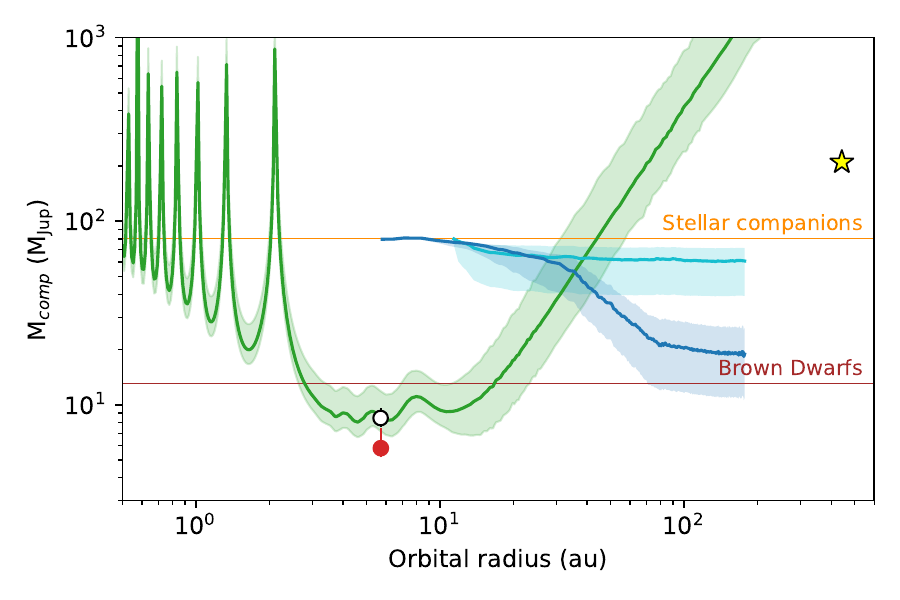}
          \caption{Proper motion anomaly sensitivity curve for HD\,57625. The dark green curve shows the PMa-compatible companion masses as a function of orbital separations, the shaded region corresponding to the 1$\sigma$ uncertainty range. The blue and cyan curves represent the SHARK-NIR and LMIRCam mass limits as in Fig.~\ref{fig:contrast}. Horizontal lines indicate the  brown dwarf (brown) and stellar mass (orange) thresholds. The positions of HD\,57625\,b as obtained by the RV-only fit and by the joint RV and PMa fit are shown as a red and white circle, respectively, while the distant stellar companion is shown as a yellow star.}
          \label{fig:pma}
        \end{figure}
        \begin{figure}
          \includegraphics[width=\linewidth]{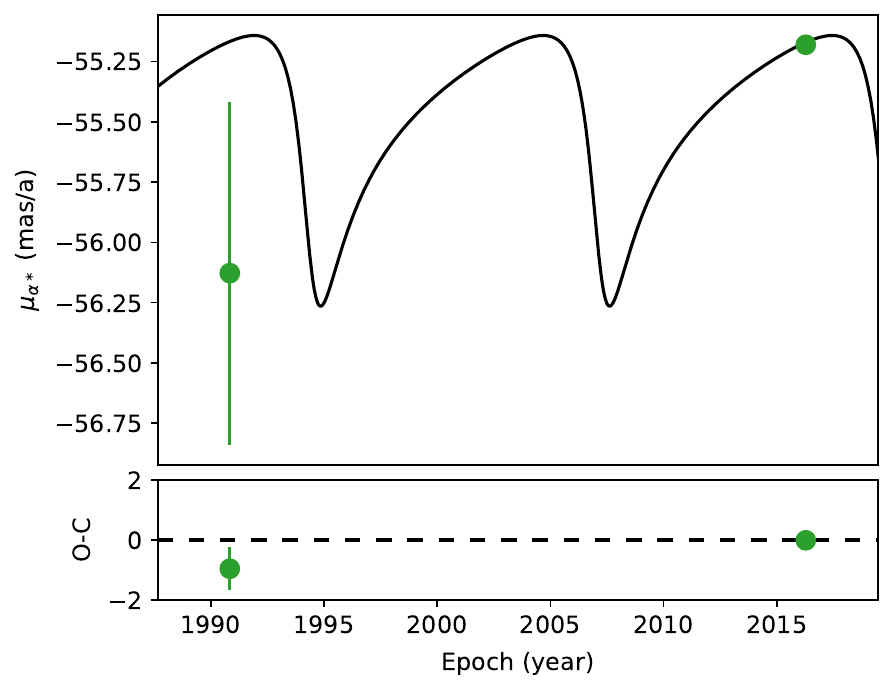}
          \\
          \includegraphics[width=\linewidth]{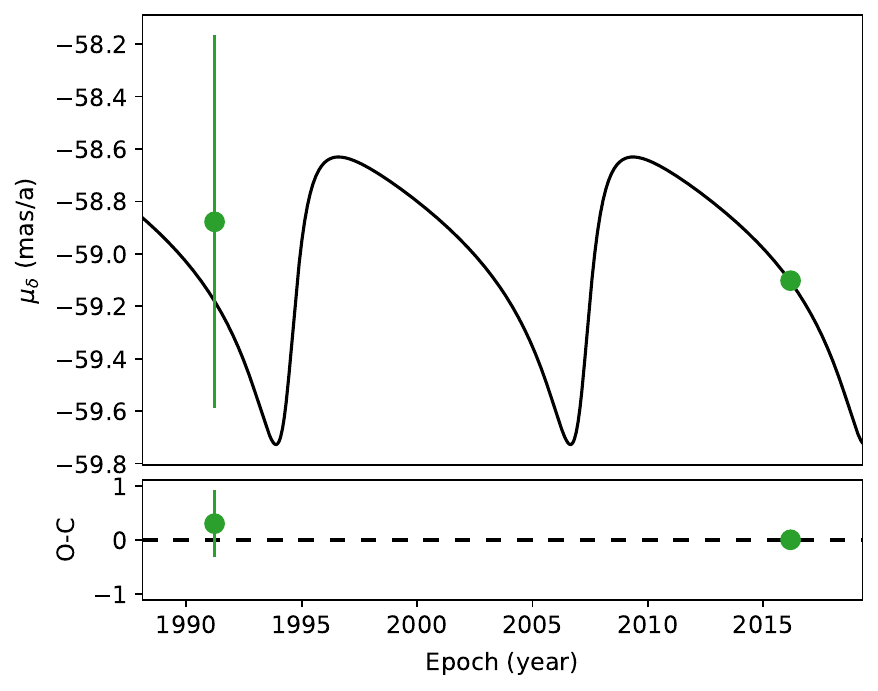}
          \caption{Observed and fitted proper motions in right ascension (\textit{top panel}) and declination (\textit{bottom panel}). The best-fit orbit obtained by the joint RV and PMa fit is shown as a black curve, with the proper motion measurements from \hip and Gaia EDR3 shown as green circles.}
          \label{fig:pmafit}
        \end{figure}
        \begin{figure}
          \includegraphics[width=\linewidth]{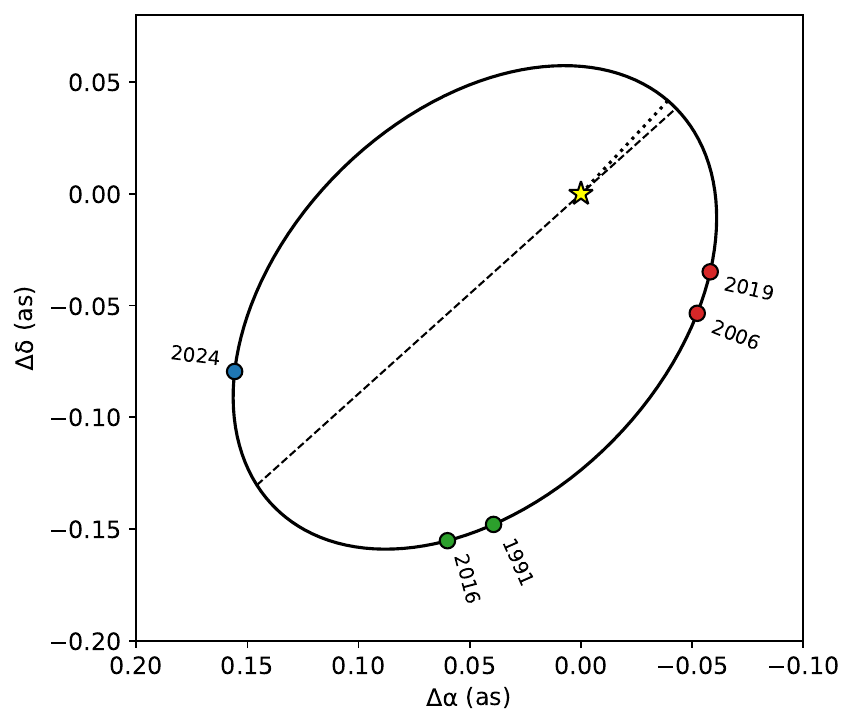}
          \caption{Relative astrometric orbit and selected predicted position for HD\,57625\,b. The position of the host star is marked by a yellow star, with a dashed line representing the line of nodes and a dotted line connecting the host star to periastron. Coloured circles mark the predicted positions of the planetary companion at the epoch of our imaging observation (2024, blue), at the starting and final epochs of available RVs (2006 and 2019, red) and at the reference epochs for \hip and \gdr{3} (1991 and 2024, green)}
          \label{fig:orbit}
        \end{figure}
        As previously discussed, one of the main hints of the presence of a companion orbiting HD\,57625 is the highly significant variation in proper motion measurements between the \hip and \gaia DR3 epochs, which \cite{kervella2022} reports as having a SNR of 11.04. Assuming a $1.16\pm0.06$\Msun mass for the host star, the catalog estimates a companion dynamical mass of 11.18\Mjup at a separation of 3\au, of 8.91\Mjup at 5\au, of 9.33\Mjup at 10\au and of 37.45\Mjup at 30\au, indicating that the observed PMa is compatible with either a planetary-mass companion at short-to-wide separations or a brown dwarf or stellar companion at larger separations.
        \par In Fig.~\ref{fig:pma}, following the method detailed in \cite{kervella2022}, the companion masses compatible with the observed astrometric acceleration is plotted as a function of possible orbital separations. We note that the distant stellar companion mentioned in Sect.~\ref{sec:star-parameters}, shown in Fig~\ref{fig:pma} as a yellow star, is not compatible with the observed PMa, as a companion with its assumed mass of $\sim$0.2\Msun would be responsible for the reported acceleration only at an orbital separation of $\sim$70\au, much closer than the projected separation of $440$\au of the known stellar companion.
        \par On the other hand, the RV-detected companion with minimum mass $5.79_{-0.63}^{+1.71}$\Mjup and semi-major axis $5.71_{-0.21}^{+0.33}$\au discussed in Sect.~\ref{sec:rv} (shown in Fig.~\ref{fig:pma} as a red circle) is compatible within 2$\sigma$ with the observed PMa. The same Figure additionally shows that only planetary-mass companions satisfactorily explain the observed PMa within the orbital separation compatible with the RV signal detected in the SOPHIE public data, strongly suggesting a planetary true mass companion to be the origin of both the RV and PMa signals.
        \par To provide a more complete characterization of the orbital elements of HD\,57625\,b, and more specifically a precise value of dynamical mass, we used the Python code \orvara \citep{brandt2021orvara}, an MCMC code designed to fit Keplerian orbits to any combination of proper motion variations, absolute and relative astrometry, and radial velocities to obtain precise dynamical masses and orbital elements. More specifically, \orvara is designed to use the PMa computed and reported in the \hip-\gaia Catalog of Accelerations \citep[HGCA,][]{brandt2018,brandt2021hgca}, in which the \hip and \egdr{3} catalogues have been cross-calibrated to account for systematics and shift all proper motions in the \egdr{3}.
        \par The best-fit orbital parameters obtained by the joint RV-PMa fit are listed in Table~\ref{table:rvpmafit}, with best-fit proper motion anomaly curves shown in Fig.~\ref{fig:pmafit} and relative astrometric orbit with predicted positions at selected epochs for the planetary companion shown in Fig.~\ref{fig:orbit}. This solution characterizes HD\,57625\,b as having an orbital period of ${4843}_{-167}^{+306}$\days, eccentricity of ${0.52}_{-0.03}^{+0.04}$, inclination of ${43.82}_{-7.22}^{+14.30}$deg and dynamical mass of ${8.43}_{-0.91}^{+1.1}$\Mjup.
        \par We show the companion's position in the mass-separation parameter space as a white circle compared with the PMa sensitivity curve in Fig.~\ref{fig:pma}, highlighting the solution's agreement with the observed PMa. We note that the orbital period derived from the joint RV-PMa solution is compatible and more precise than the RV-only one obtained in Sect.~\ref{sec:rv}, and the same applies to the companion dynamical mass which additionally lies firmly below the traditionally assumed 13\Mjup deuterium-burning threshold, confirming the planetary nature of HD\,57625\,b.
        \begin{table}
            \caption{Joint RV and PMa best-fit orbital solution for HD\,57625\,b.}    \label{table:rvpmafit}
            \centering
            \begin{tabular}{l c c}
                \hline\hline
                    Parameter & Priors & Best-fit values\\
                \hline
                    $M_b$ (\Mjup)& $1/M$ (log-flat) &${8.43}_{-0.91}^{+1.10}$ \\[3pt]
                    $a_{\rm b}$ (au) & $1/a$ (log-flat) &${5.70}_{-0.13}^{+0.14}$ \\[3pt]
                    \sresinob & \uprior$(-1,1)$ &${0.008}_{-0.056}^{+0.068}$ \\[3pt]
                    \srecosob & \uprior$(-1,1)$ &${0.720}_{-0.026}^{+0.023}$ \\[3pt]
                    $i_{\rm b}$ (deg) & $\cos{i}$, \uprior$(0,180)$ &${43.82}_{-7.22}^{+14.30}$ \\[3pt]
                    $\lambda_0$ (deg) & \uprior$(0,360)$ &${89.4}_{-6.1}^{+7.1}$ \\[3pt]
                    $\Omega_{\rm b}$ (deg) & \uprior$(0,360)$ &${301}_{-40}^{+13}$ \\[3pt]
                    $P_{\rm b}$ (\days) & derived &${4843}_{-167}^{+306}$ \\[3pt]
                    $\omega_{\rm b}$ (deg) & derived &${9.2}_{-7.2}^{+349}$ \\[3pt]
                    $e_{\rm b}$ & derived &${0.52}_{-0.03}^{+0.04}$ \\[3pt]
                    $j_{\rm SOPHIE}$ (\mps) & $1/j$ (log-flat) &${8.8}_{-2.0}^{+2.9}$ \\[3pt]
                    $j_{\rm SOPHIE+}$ (\mps) & $1/j$ (log-flat) &${6.76}_{-0.68}^{+0.78}$ \\[3pt]
                \hline
            \end{tabular}
        \end{table}
        
    \section{Detection completeness}    \label{sec:completeness}
        \begin{figure*}
          \includegraphics[width=0.49\linewidth]{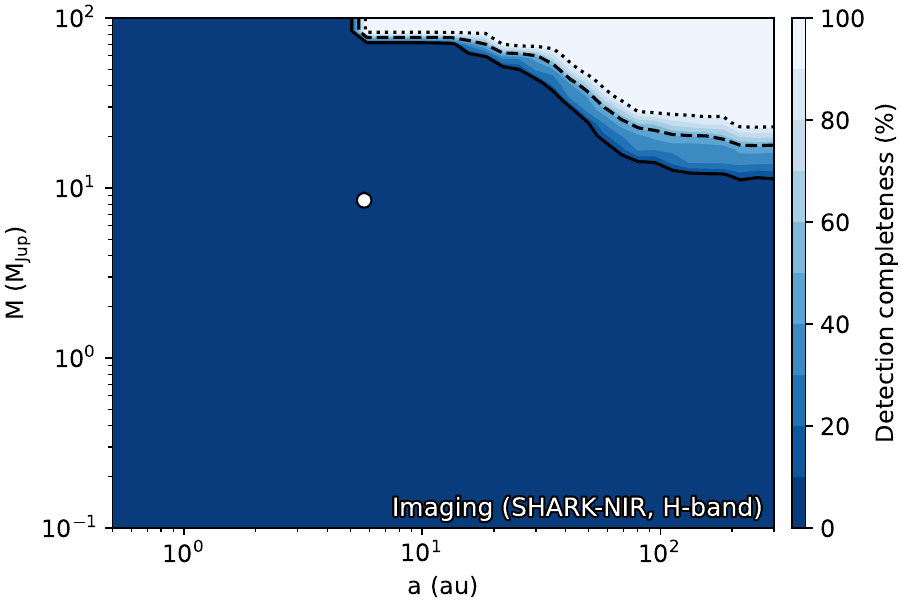}
          \includegraphics[width=0.49\linewidth]{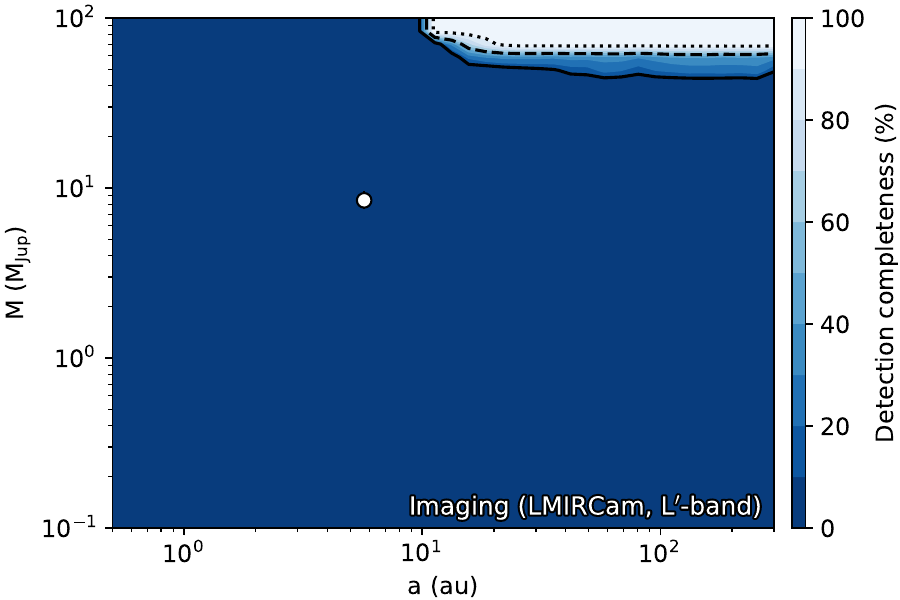}
          \\
          \includegraphics[width=0.49\linewidth]{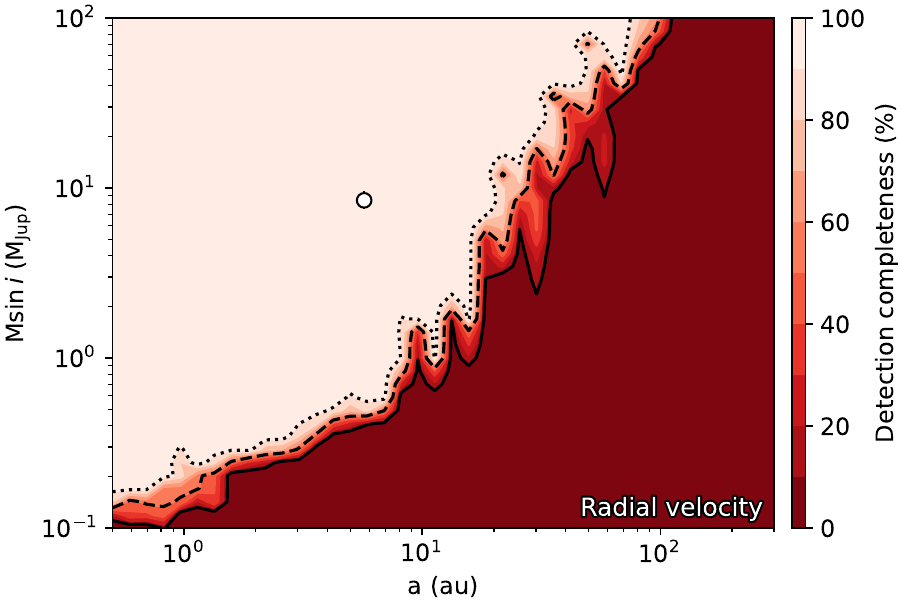}
          \includegraphics[width=0.49\linewidth]{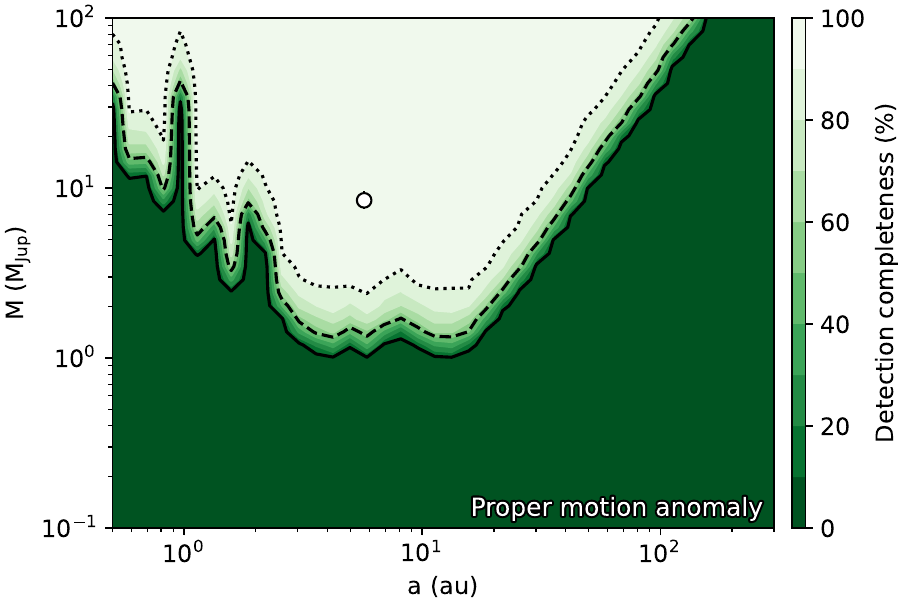}
          \\
          \centering
          \includegraphics[width=0.49\linewidth]{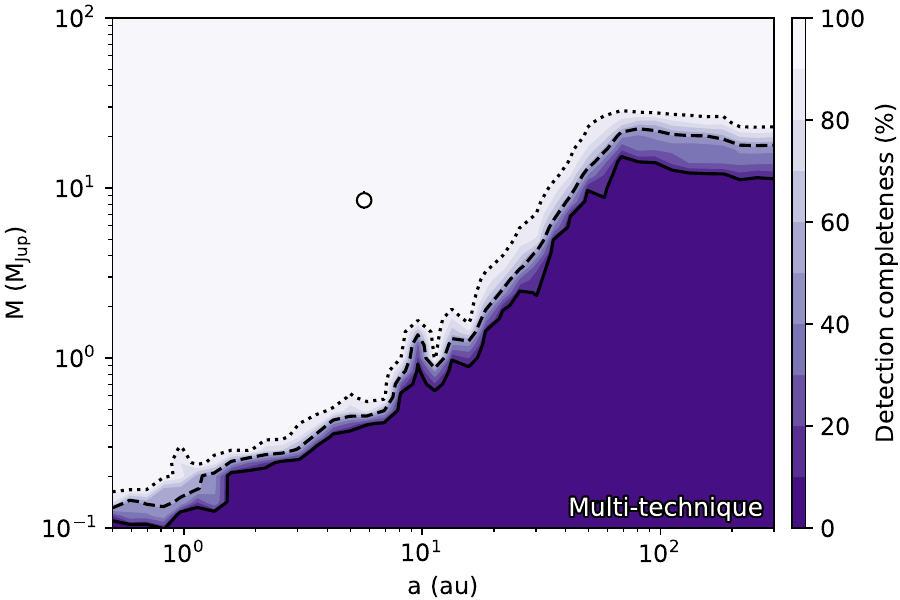}
          \caption{Detection completeness maps for the imaging (blue, \textit{top row}), radial velocity (red, \textit{middle left panel}), proper motion anomaly data (green, \textit{middle right panel}) discussed in this work, as well as the multi-technique global map (purple, \textit{bottom panel}). In all panels, the solid, dashed and dotted curves identify the 10\%, 50\% and 100\% detection completeness thresholds, a white circle marks the position of the detected planetary companion as obtained by the joint RV and PMa fit.}
          \label{fig:detection-completeness}
        \end{figure*}
        In addition to allowing the detection of the planetary companion HD\,57625\,b, the imaging, RV and PMa data analysed in the previous Sections can be further used to assess the detection completeness of the datasets. By doing so, we can identify the full-completeness regions in the mass-separation parameter space in which the available data can exclude the presence of additional companions or, on the contrary, the low-completeness regions in which additional companions could, in principle, be detected by follow-up observations.
        \par In order to do so, we follow an injection and retrieval scheme in which synthetic companions are injected into the available datasets and subject to statistical tests to assess their detectability. We use a Python code we developed to assess the detection completeness of RV time series analysed in previous works \citep{barbato2018,barbato2019,barbato2023,matthews2023,gratton2024,mesa2024}, now expanded to also be used on imaging and astrometric data in order to produce multi-technique detection completeness maps.
        \par We explore a grid of 40$\times$40 companion masses and orbital separations, respectively ranging from 0.1 to 100\Mjup (considering true mass $M$ for imaging and PMa data, and minimum masses $M\sin{i}$ for RV data) and from 0.5 to 300\au. This wide parameter space spans a large variety of companion categories, from close-in giant planets to wide-orbit low-mass stellar companions, allowing both the investigation of different possible companions and the showcasing of each datasets discovery space and synergic interplay. For each mass-separation realization, we generate $10^3$ sets of randomly drawn values of $e$, $\omega$, $\Omega$, $\lambda_0$, $i$ and stellar age used to compute as many synthetic H and L$^\prime$-band contrast, RV time series and PMa signals corresponding to the injected companion. We note that the age values are drawn from a uniform distribution between the lower and upper values of stellar age derived for HD\,57625 in Sect.~\ref{sec:star-parameters}, in order to account for this parameter uncertainty and the dependence of H-band injected companion emission we compute using the AMES-COND models, and therefore of the detecting capability of imaging observations, as a function of the system age.
        \par We then perform a statistical test to assess the detectability of each of the $1.6\cdot10^6$ synthetic companions with each of the considered techniques, considering the injected signal detected with imaging if the computed contrast is higher than the measured non-detection contrast curve at the corresponding separation (see Fig.~\ref{fig:contrast}), with radial velocity if the injected companion produces a periodogram peak with FAP$\leq10^{-3}$ and by astrometry if the injected proper motion anomaly has SNR$\geq3$. Finally, for each mass-separation realization we compute the detection completeness as the ratio between detectable injected signals and total injections. The detection completeness maps thusly obtained for all techniques considered in this work are shown in Fig.~\ref{fig:detection-completeness}, the position of the detected exoplanet marked by a white circle.
        \par Starting with imaging (top row), it is immediate to note that the detection completeness of our SHARK-NIR H-band observation (left panel) is limited for lower orbital separations by its IWA of 120\mas, corresponding to 5.30\au given the nominal stellar distance of 44.24\pc. Above this lower limit on separation, we achieve full completeness only for stellar companions (>80\Mjup) at all separations explored, and for massive brown dwarfs (>30\Mjup) beyond 80\au. Our LMIRCam L$^\prime$-band observation (right panel) is found to be more limited for close-in companions, due to its higher IWA of 246\mas corresponding to 10.88\au for our star, and provides full completeness only for outer stellar (>80\Mjup) companions. While the planetary companion detected with radial velocities and proper motion anomaly remains firmly beyond the detection capabilities of both SHARK-NIR and LMIRCam, by virtue of its relatively low mass and orbital separation comparable and lower than SHARK-NIR and LMIRCam IWA respectively, our observations allow to robustly exclude the presence of massive brown dwarfs or stellar companions at all separations of interest, once again stressing the planetary nature of the detected RV and PMa signal. Additionally, we note the presence of significant discovery space for wide orbit giant planetary companions for follow-up deep imaging observations.
        \par Considering instead radial velocity (middle left panel), the public SOPHIE data provide full detection completeness above minimum masses of 0.2\Mjup within 1\au, above 1.5\Mjup within 10\au and finally above 20\Mjup below 30\au, allowing us to exclude the presence of additional companions in the system in these regions in the parameter space. On the other hand, the available RV data is still blind to the presence of less massive giant exoplanets on wide orbits, such as companions with $M\sin{i}<0.2$\Mjup beyond 1.5\au, <1\Mjup beyond 10\au and <10\Mjup beyond 40\au. Therefore, follow-up RV observations still have some significant discovery space available for the detection of long-period giant planets, as well as lower-mass companions on a large variety of orbits.
        \par Finally, the astrometry detection map (middle right panel) shows that the \hip-\gaia proper motion anomalies precision provide full completeness for massive companions, such as those with true mass larger than 4\Mjup between 2.5 and 15\au, as well as larger than 30\Mjup below 50\au, allowing us to further exclude the presence of additional massive companions in the system within these orbital ranges. On the other hand, a significant portion of giant planet mass regime remains below the current detection capabilities of the PMa measurements for HD\,57625, with no sub-Jovian companion being detectable at any of the orbital separations explored, allowing for significant exoplanetary discovery space with full astrometric orbit measurements in future Gaia releases.
        \par To conclude, we additionally shown in the bottom panel of Fig.~\ref{fig:detection-completeness} a multi-technique global detection completeness map, obtained by considering each injected companion detected by at least one of the datasets studied so far. From this map is evident how the multi-technique observations of HD\,57625 allow to compensate each technique's weaknesses and provide a rather complete view of the system, with RV data probing the inner, low-mass region of the parameter space to which the other techniques are blind, PMa providing completeness for intermediate-to-large separation giant companions and imaging probing the more massive and outer regions of the system, once again highlighting the importance of multi-technique observations and analysis for the characterization of planetary systems.
        
    \section{Discussion and conclusions} \label{sec:conclusions}
        In this work we reported the detection and characterization of HD\,57625\,b, the giant planetary companion responsible for the astrometric acceleration of its host star, via a synergic analysis of imaging, radial velocity and astrometric datasets. To investigate the observed significant proper motion anomaly between Hipparcos and \gdr{3}, we performed high-contrast H-band observations with SHARK-NIR, the near-infrared camera recently installed at the Large Binocular Telescope, in parallel with LMIRCam in L$^\prime$, resulting in a non-detection that allowed us to confirm a substellar nature for the unseen companion causing the observed acceleration.
        \par By jointly analysing public archival SOPHIE radial velocity time series and the Hipparcos-\gdr{3} proper motion anomaly measurements we further confirm the planetary nature of the companion, characterize its orbit as well as determining its true mass even facing incomplete orbital coverage by the available radial velocity measurements. We find HD\,57625\,b to be a giant exoplanet with a true mass of ${8.43}_{-0.91}^{+1.1}$\Mjup on a ${5.70}_{-0.13}^{+0.14}$\au orbit with a relatively high eccentricity of ${0.52}_{-0.03}^{+0.04}$. To further analyse the planetary system and fuel future observations, we additionally performed a multi-technique detection completeness assessment, finding significant yet-unexplored discovery space for additional outer giant and inner low-mass companions, as well as underlying the importance of synergic multi-technique analysis in fully characterizing exoplanetary systems.
        \par A noteworthy characteristic of HD\,57625\,b is its moderate-to-high value of orbital eccentricity, which for giant planets usually interpreted as the result of either Kozai-Lidov effects \citep{kozai1962,innanen1997,wu2003}, in which the planetary eccentricity growth is driven by secular interactions with a distant stellar or brown dwarf companion, or planet-planet scattering \citep{weidenschilling1996,chambers1996,raymond2009}, in which gravitational instability in multi-planet systems leads to repeated orbital encounters often resulting in multiple surviving planetary companions with high eccentricities. 
        \par While our multi-technique detection completeness assessment proves that yet-undiscovered outer giant planetary companions could still be present in the HD\,57625 system, the lack of inner massive companions and especially the presence of the wide-orbit stellar companion 2MASS J07251770+5634002 (\gdr{3} 989016790859521280) suggest a Kozai-driven origin for the observed eccentricity of HD\,57625\,b. Indeed, following \cite{ford2000} and \cite{takeda2005} we can estimate the timescale for Kozai-driven eccentricity modulation as:
        \begin{equation}    \label{eq:kozai-timescale}
            P_{\rm Kozai} \simeq P_{\rm b}\frac{M_{\rm A}+M_{\rm b}}{M_{\rm B}} \left( \frac{a_{\rm B}}{a_{\rm b}} \right)^3 \left( 1-{e_{\rm B}}^2 \right)^{3/2}
        \end{equation}
        with the indexes b, A and B referring to the planetary companion, the primary and secondary component of the binary system, respectively. As we have estimates of $M_{\rm B}\sim0.2$\Msun and $a_{\rm B}\sim440$\au for the distant stellar companion but no assessment on its orbital eccentricity, we consider extreme values of 0 and 0.9 for $e_{\rm B}$. Accounting for the uncertainties in host star and planetary companion parameters, we obtain values of $P_{\rm Kozai}$ ranging from 26 to 39 Ma assuming $e_{\rm B}=0$ and from 2 to 3 Ma for $e_{\rm B}=0.9$. As these timescales are considerably lower than the host star age of $4.8_{-2.9}^{+3.7}$ Ga, it is clear that the planetary system has undergone a large number of oscillation cycles, and as such we identify Kozai interaction as a likely responsible for the planetary companion eccentricity. However, in the absence of information on the stellar companion orbital orientation, we are currently unable to determine whether the the planet-secondary relative orbital inclination is higher than the threshold value of 39.2\fdg needed to incite Kozai oscillations. While in principle the combination of proper motion vector and radial velocity measurements for the wide-orbit stellar companion can be used to provide a characterization of its orbital inclination, no value of radial velocity is available in \gdr{3} or in previous literature. Furthermore, the stellar companion is not a part of the Hipparcos catalog, preventing any orbital characterization using PMa. As such, the relative orbital inclination of HD\,57625\,b and 2MASS J07251770+5634002 is impossible to determine, although future Gaia DRs could provide additional measurements and help shed light on the system dynamical history.
        \par Finally, we highlight that HD\,57625\,b joins the population of long-period giant planetary companions having true mass value determined via the synergic usage of multiple detection methods. As of the time of writing, only a few tens of such exoplanets are known, most of which have had their dynamical mass characterized only in recent years thanks to the usage of Gaia astrometric measurements and the growing multi-technique analysis approach in exoplanetology. As such, many studies assessing exoplanetary occurrence rates often characterize detected massive companions based on minimum mass estimates or the presence of long-term RV trends \citep[see e.g.][]{bryan2019,rosenthal2022} and therefore are highly dependant on the inherent planetary mass degeneracy in the absence of additional constraints. Therefore, this small but growing sample of outer giant exoplanets with true mass determination represents an important asset in exoplanetology, and the growing multi-technique exoplanetology approach as well as the future Gaia data releases and astrometric solutions will continue to prove essential in exploring the observed variety of exoplanetary system architectures, as well as furthering our understanding of the formation and dynamical evolution processes that shaped such variety.
    
    \begin{acknowledgements}
        The authors wish to thank the anonymous referee for their useful comments.
        We thank Tom Herbst from MPIA-Heidelberg and the LINC-NIRVANA team for sharing part of their instrument control SW to operate the motorized axis of SHARK-NIR. 
        We also express our appreciation to NASA and Marcia Rieke, the Principal Investigator of JWST/NIRCam, for granting us the opportunity to utilize one of the NIRCam spare detectors as the primary detector for the SHARK-NIR scientific camera. 
        Observations have benefited from the use of ALTA Center (\url{alta.arcetri.inaf.it}) forecasts performed with the Astro-Meso-Nh model. Initialization data of the ALTA automatic forecast system come from the General Circulation Model (HRES) of the European Centre for Medium Range Weather Forecasts.
        The LBT is an international collaboration among institutions in the United States, Italy and Germany. The LBT Corporation partners are: The University of Arizona on behalf of the Arizona university system; Istituto Nazionale di Astrofisica, Italy; LBT Beteiligungsgesellschaft, Germany, representing the Max Planck Society, the Astrophysical Institute Potsdam, and Heidelberg University; The Ohio State University; The Research Corporation, on behalf of The University of Notre Dame, University of Minnesota and University of Virginia. 
        DB and AR wish to thank J. Gomes da Silva for his great availability and invaluable support in adapting and applying \texttt{ACTIN2} to this work.
        AR acknowledges support by the Fondazione ICSC , Spoke 3 Astrophysics and Cosmos Observations. National Recovery and Resilience Plan (Piano Nazionale di Ripresa e Resilienza, PNRR) Project ID CN\_00000013 "Italian Research Center on High-Performance Computing, Big Data and Quantum Computing"  funded by MUR Missione 4 Componente 2 Investimento 1.4: Potenziamento strutture di ricerca e creazione di "campioni nazionali di R\&S (M4C2-19)" - Next Generation EU (NGEU).
        This work has made use of data retrieved from the SOPHIE archive at Observatoire de Haute-Provence (OHP), available at \url{atlas.obs-hp.fr/sophie}.
        This work has made use of data from the European Space Agency (ESA) mission {\it Gaia} (\url{https://www.cosmos.esa.int/gaia}), processed by the {\it Gaia} Data Processing and Analysis Consortium (DPAC, \url{https://www.cosmos.esa.int/web/gaia/dpac/consortium}). Funding for the DPAC has been provided by national institutions, in particular, the institutions participating in the {\it Gaia} Multilateral Agreement. 
        This research has made extensive use of the NASA-ADS, SIMBAD and Vizier databases, operated at CDS, Strasbourg, France.
    \end{acknowledgements}
  
  \bibliographystyle{aa}
  \bibliography{ref}
  
\end{document}